\documentstyle[aps,array]{revtex}
\def\s#1{\slash\!\!\!{#1}}

\def\bee{\begin{eqnarray}}
\def\eee{\end{eqnarray}}
\def\nn{\nonumber\\}

\def\fr#1#2{\frac{#1}{#2}}

\begin{document}
\draft
\title{The Applicability of PQCD and NLO Power Corrections for Pion Form Factor}
\author{Tsung-Wen Yeh\thanks{E-mail: twyeh@cc.nctu.edu.tw}}
\address{Institute of Physics, National Chiao-Tung University,
Hsinchu, Taiwan 300, Republic of China}
\maketitle
\begin{abstract}
In many years ago, Isgur and Llewellyn Smith addressed that PQCD is inapplicable to exclusive processes \cite{Isgur:1984jm,Isgur:1989iw}, such as the pion form factor. The main problem  is that the asymptotic of PQCD is only about one fourth of the experimental value \cite{Bebek:1976wv,Bebek:1978pe,Amendolia:1986wj}. We reexamine this PQCD deep problem. By including NLO power corrections to the pion form factor, we may arrive at a perturbative explanation for the data. The key realization is that we need to interprete that the strong interaction coupling constant involved in the PQCD result should be taken as an effective coupling constant under nonperturbative QCD vaccum. This implies that one can equally identify the relevant scale for the effective coupling constant as the factorization scale about 1 GeV. We also find that the average momentum fraction variable locates about 0.5, which is in favor of the asymptotic pion wave function. By employing photon-pion form factor with NLO power corrections to factorize out the nonperturbative effects involved in the effective coupling constant, we can extract an effective running coupling constant, which represents an effective coupling involving in the hard scattering subprocesses. The difference between the effective running coupling constant and the usual perturbative running coupling constant ($\Lambda_{\makebox{\small QCD}}=0.3$ GeV) is very small for $Q^2> 1$ GeV$^2$.  The effective running coupling constant $\alpha_s/\pi$ is smaller than $0.2$ for $Q^2>1$ GeV$^2$.  This directly showes that PQCD is applicable to exclusive processes at energy $Q^2> 1$ GeV$^2$. In summary, with NLO power corrections, PQCD can completely explain the $Q^2$ spectrum of pion form factor.
\end{abstract}

\vskip 2.0cm

\section{INTRODUCTION}
The exclusive process plays an important role in understanding the strong interactions. The detail analysis of exclusive processes not only can reveal the constituents of hadron but also can untangle the underlying dynamics. One important progress in this respect is the prediction of perturbative QCD (PQCD) for exclusive process in high energy limit \cite{Lepage:1979zb,Lepage:1980fj}. In PQCD formula, the transition amplitude can be expressed as a convolution integral over initial and final state meson wave fucntions and a hard funtion. The hard function can be perturbatively calculable, while the meson wave functions can only be determined, nonperturbatively. 

The pion form factor has been investigated in the framework of PQCD \cite{Lepage:1979zb,Lepage:1980fj,Li:1992nu,Field:1981wx,Melic:1999qr,Efremov:1980qk,Geshkenbein:1982zs,Geshkenbein:1982zu,Geshkenbein:1984qn,Chernyak:1984ej,Jakob:1993iw,Cao:1999st}. In the experimentally accessable energy region of few GeV$^2$, the asymptotic of PQCD is only about one fourth of the experimental value. This leads to two problems for PQCD. One problem relates to the self-consistency of PQCD and the other problem requires increasing the maginutde of PQCD. The former was first addressed by Isgur and Llywellyn Smith \cite{Isgur:1984jm,Isgur:1989iw} that the most contributions come from the soft end-point regions or small values of the fraction variables, indicating the perturbative calculation inconsistent. This problem was partially resolved by Li and Sterman \cite{Li:1992nu} who proposed a modified PQCD formulation, in which the physics of the end-point regions are carefully dealt with and the associated contributions are suppressed by a resummation over soft radiative corrections, the Sudakov factor. It was found that, with the transverse degrees of freedom playing the role of infrared cut-off, the PQCD contribution becomes self-consistent for momentum transfers as low as few GeV. However, note that the magnitude of modified PQCD is still smaller than the experimental value by a factor of $2\sim 4$.

The general wisdom to increase the magnitude of PQCD is by invoking the power correction \footnote{The contributions come from the NLO in $\alpha_s$ is $20\%\sim 30\%$ \cite{Field:1981wx,Melic:1999qr}.}. However, the conclusion of past results was rather pessimistic. Even including twist-3 power corrections, it is still difficult to accomodate the data. More worsely, introducing the twist-3 contribtion imposes another problem that the twist-3 contributions may exceed the leading twist contribution at intermediate energy region of $Q^2$ being $2\sim 15$ GeV$^2$\cite{Geshkenbein:1982zs,Geshkenbein:1982zu,Geshkenbein:1984qn} or $2\sim 40$ GeV$^2$ (for the parton carrying transverse momenta) \cite{Cao:1999st}.  This is because there involves a huge factor  $m_{\pi}^2/m_0\approx 1000$ for the twist-3 contribution, where $m_{\pi}$ is the pion mass and $m_0$ is the average constituent quark masss of the pion. The strange behavior of the twist-3 contribution violates the principle of perturbation theory: the magnitude of subleading terms should be smaller than that of the leading ones.  

The reason why the twist-3 contribution is inconsistent with perturbation theory may reside in the method of calculation. In the previous approaches \cite{Geshkenbein:1982zs,Geshkenbein:1982zu,Geshkenbein:1984qn,Cao:1999st}, the spin projections $\gamma_5$ and $\gamma_5\sigma^{\mu\nu}$ are employed and the normalizations of related distribution amplitudes (DAs) are determined with the help of chiral perturbation theory. However, these methods gave no garantees on factorization of hard and soft radiative corrections for a twist-3 contribution. The failure of these methods in explaining the data may indicate the incorrectness of factorization.  

In a recent power expansion method developed by us \cite{Yeh:2001gu}, it showes that the calculation of high twist contribution is compatible with QCD factorization. There requires more perturbative contributions than spin projections. The most important feature of this method is that the partonic interpretation for high twist contribtion is retained. For the pion form factor, the NLO power corrections calculated by this method have reverse signs and inverse magitudes against to those results of the previous mentioned methods. These differences give us a opportunity to explain the data by only employing PQCD. It is noted that the NLO contribution derived in this text is smaller than the leading order contribution over whole energy range of the experiment.

Our organization is as follows. We investigate power expansion and QCD factorization for process $\gamma^*\pi\to\pi$ in Sec.~II. This is a preliminary for Sec.~III. The pion form factor $F_\pi(Q^2)$ upto order of $O(Q^{-4})$ are evaluated in Sec.~III. The proof showing that PQCD is applicable to pion form factor is present in here. A discussion for the reason why Isgur and Llewellyn Smith made their conclusion is explained in Sec.~IV. Sec.~V is devoted to conclusions.

\section{Collinear Expansion and QCD FACTORIZATION}
In this section, we will describe our approach of power expansion for $\gamma^*\pi\to\pi$. The method we employ is called the collinear expansion \cite{Yeh:2001gu,Ellis:1982wd,Qiu:1990dn}. In the following, we first sketch the procedures of how to perfom the collinear expansion for $\gamma^*\pi\to\pi$. Then, we will show the collinear expansion is compatible with the QCD factorization, which is important for giving a reliable perturbative calculation in QCD.

\subsection{Tree Level Collinear Expansion}
Let $\sigma=\phi^*(P_2,k_2)\otimes\sigma_p(k_1,k_2)\otimes\phi(P_1,k_1)$ represent the lowest order amplitude for $\gamma^*(q)\pi(P_1)\to\pi(P_2)$ as depicted in Fig.~1(a).  The $\sigma_p(k_1,k_2)$ denotes the amplitude for partonic subprocess and the $\phi(k_i)$, $i=1,2$, represents the pion DAs. The $\otimes$ means convolution integral over the loop momenta $k_i$. To pick out the leading contribution, we assign the momenta of the initial and final state pions in the following way. We let $P_i=Q v_i$ where the dimensionless vectors obey $v_i^2=0$, $v_1\cdot v_2=\frac{1}{2}(1-\cos{\theta})$. $\theta$ is the center of mass scattering angle.
The internal loop momenta $k_i^{\mu}$, $i=1,2$ are parameterized as
\begin{eqnarray}
k_i^{\mu}=x_i Q v_i^\mu + y_i v_i^{\prime}+k_{i\perp}
\end{eqnarray}
$x_i$ are dimensionless number of order unit, while $y_i$ have dimensions of mass. The vector $v_i^{\prime}$ are in the direction of the opposite-moving external vectors such that $v_i\cdot v_i^{\prime}=1$, $(v_i^{\prime})^2=0$ and
$v_1^{\prime}\cdot v_2^{\prime}=v_1\cdot v_2$, and also $\epsilon_{ij}v_i\cdot v_j^{\prime}=\frac{1}{2}(1+\cos\theta)$ for $i,j=1,2$. $y_i$ are solved by keeping $k_i^2$ invariant
\begin{eqnarray}
y_i=\frac{k_i^2+k_{i\perp}^2}{2x_i Q}\ .
\end{eqnarray}
The first step is to perform Taylor expansion for the parton amplitude 
\bee\label{st1}
\sigma_p(k_i)=\overline{\sigma}_p(k_i=x_i P_i)
                  +{(\overline{\sigma}_p)_{\alpha}(x_i,x_i) }w^\alpha_{i\alpha^\prime}k^{\alpha^\prime}_i+\cdots
\eee
where we have assumed the low energy theorem 
\bee
\fr{\partial}{\partial k^\alpha}_i\sigma_p(k_i)\Bigg|_{k_i=x_i  P_i}=(\overline{\sigma}_p)_{\alpha}(x_i,x_i)\ . 
\eee  
and have employed $w^\alpha_{i\alpha^\prime}k^{\alpha^\prime}_i=(k_i-x_i P_i)^\alpha$ and $w^\alpha_{i\alpha^\prime}=g^\alpha_{\alpha^\prime}-v^\alpha_i v_{i\alpha^\prime}^{\prime}$.
The leading term $\phi^*\otimes \overline{\sigma}_p\otimes \phi$ containes both leading-twist (LT) and next-to-leading-twist (NLT) contributions in accord with the structures of $\overline{\sigma}_p$,
the terms proportional to $\s{v}^{\prime}_i$ or $\s{v}_i$. The $\s{v}^{\prime}_i$ terms would project collinear $q\bar{q}$ pair from the meson, while $\s{v}_i$ terms would not diminish only when the $q\bar{q}$ pair carry noncollinear momenta. The second step is to substitute the leading parton amplitude $\overline{\sigma}_p$ into the convolution integral with the meson wave function $\phi$ to extract high twist contributions 
    \bee\label{st2}
     \phi^*\otimes \overline{\sigma}_p\otimes \phi=\phi^*_0\otimes \overline{\sigma}_p\otimes\phi_0 +
     \phi^*_0\otimes \overline{\sigma}_p\otimes\phi_1 +\phi^*_1\otimes \overline{\sigma}_p\otimes\phi_0 + \cdots ,
    \eee
where $\phi_0$ and $\phi_1$ denote the leading and next-to-leading meson DAs, respectively. The high twist DA $\phi_1$ contains both short distance and long distance contributions. The short distance contributions of $\phi_1$ arise from the noncollinear components of $k_i$. By the equations of motion, the noncollinear components of $k_i$ will induce one quark-gluon vertex $i\gamma_\alpha$ and one special propagator $i\s{v}^{\prime}_i/2 x_i Q$ \cite{Qiu:1990dn}. Because that the special propagator is not propagating, the quark-gluon vertex and the special propagator should be included into the hard function, $\overline{\sigma}_p$. In this way, we may factorize $\phi_1$ as $\phi_1\approx(\phi_1^H)_\alpha w^\alpha_{\alpha^\prime}(\phi_1^S)^{\alpha^\prime}$ and absorb the short distance piece $(\phi_1^H)_\alpha$ into $\overline{\sigma}_p$. It leads to the third step 
%  \item Step 3:
\bee\label{st3}
     \phi^*_0\otimes \overline{\sigma}_p\otimes\phi_1&=&\phi^*_0\otimes \overline{\sigma}_p
        \otimes((\phi_1^H)_\alpha\bullet w^\alpha_{\alpha^\prime} (\phi_1^S)^{\alpha^\prime})\nn
      &=&\phi^*_0\otimes (\overline{\sigma}_p\bullet\phi_1^H)_\alpha\otimes w^\alpha_{\alpha^\prime} (\phi_1^S)^{\alpha^\prime}\ ,
\eee
where  $(\phi_1^S)^{\alpha^\prime}$ containing covariant derivative $D^{\alpha^\prime}=i\partial^{\alpha^\prime}-g A^{\alpha^\prime}$ is implied.
Notice that the light-cone gauge $v_{i}^{\prime}\cdot A=0$ assures  $w^\alpha_{i\alpha^\prime}A^{\alpha^\prime}=A^\alpha$. This is legal since each part of Eq.~(\ref{st3}) is separately gauge invariant.
The second term of Eq.(\ref{st1}) also contribute to high twist corrections, as it convolutes with $\phi_0$. The momentum factor $k^\alpha$ will be absorbed by $\phi_0$ to become a coordinate derivative, denoted as  $k^\alpha\phi_0\equiv \phi_{1,\partial}^\alpha$. Consider another NLT contributions $\sigma_1\approx\phi^*_0\otimes (\overline{\sigma}_p)_\alpha\otimes w^\alpha_{\alpha^\prime}\phi_{1,A}^{\alpha^\prime}$ from Fig.~1(b) and (c), where $\phi_{1,A}^{\alpha}$ containes gauge fields. We have employed the approximation that $(\overline{\sigma}_p)_\alpha\otimes \phi_{1,A}^{\alpha}$ is the leading term of $\sigma_1$. This comes to the fourth step: 
\bee\label{st4}
     \phi^*_0\otimes (\overline{\sigma}_p)_{\alpha}\otimes w^\alpha_{\alpha^\prime} \phi_{1,\partial}^{\alpha^\prime}
+ \phi^*_0\otimes (\overline{\sigma}_p)_\alpha\otimes w^\alpha_{\alpha^\prime}\phi_{1,A}^{\alpha\prime}\equiv \phi^*_0\otimes (\overline{\sigma}_p)_1\otimes (\phi_1^S)\ ,
\eee
where we have employed $\phi_{1,\partial}^{\alpha}+\phi_{1,A}^\alpha=(\phi_1^S)^\alpha$. However, it will be found that  $(\overline{\sigma}_p)_1$ diminishes as convoluting with twist-4 DA $\phi_1^\Gamma$ (see below definition).  
Up to NLO in $Q^{-2}$, we may drop the $(\overline{\sigma}_p)_1$ term and arrive at the factorization for tree amplitudes
\bee
\sigma_0 + \sigma_1\approx \phi^*_0\otimes \overline{\sigma}_p\otimes\phi_0 +\phi^*_0\otimes (\overline{\sigma}_p\bullet\phi_1^H)\otimes \phi_1 
+\phi^*_1\otimes ((\phi_1^{*H})\bullet\overline{\sigma}_p)\otimes \phi_0
\eee
where $\phi_1$ means $\phi_1^S$.
There involves only one NLT DA $\phi_1$ for NLO power corrections.

To proceed, we need to consider the factorizations of the spin indices, the color indices and the momentum integrals over loop partons. For factorization of spin indices, we employ the expansion of the meson DA into its spin components as
\bee
\phi_{0,1} &=&\sum_{\Gamma}\phi_{0,1}^\Gamma \Gamma
\eee
where $\Gamma$ means Dirac matrix $\Gamma=1,\gamma^\mu,\gamma^\mu\gamma_5,\sigma^{\mu\nu}$. 
The factorization of the color indices take the convention that the color indices of the parton amplitudes are extract and attributed to the meson DAs. The factorization of the momentum integral is performed by making use of the fact that the leading parton amplitudes depend only on the momentum fraction variables $x_i$. The indentity can always be used
\bee
\int^1_0 dx_i\delta(x_i-k_i\cdot v_{i}^{\prime})=1\ .
\eee

The choice of the lowest twist components $\phi_{0,1}^\Gamma$ of $\phi_{0,1}$ is made by employing the power counting. 
Consider $\pi$ meson whose high twist DA $\phi^{\mu_1\cdots\mu_F;\alpha_1\cdots\alpha_B}$ has the fermion index $F$ and the boson index $B$. The fermion index $F$ arise from the spin index factorization for $2F$ fermion lines connecting DA and parton amplitude and the boson index $B$ denotes the $n_D$ power of  momenta in previous collinear expansion and the $n_G$ gluon lines as $B=n_D+n_G$. We may write
\bee\label{pc1}
\phi^{\mu_1\cdots\mu_F;\alpha_1\cdots\alpha_B}=\sum_i \Lambda^{\tau_i-1}e^{\mu_1\cdots\mu_F;\alpha_1\cdots\alpha_B}_i\phi^i
\eee
where $\Lambda$ denotes a small scale associated with DA. Spin polarizors $e_i$ denote the combinnation of  vectors $v^\mu_i$, $v^{\prime\mu}_i$ and $\gamma^\mu_\perp$. Variable $\tau_i$ represents the twist of DA $\phi^i$. The restrictions over projector $e^{\mu_1\cdots\mu_F;\alpha_1\cdots\alpha_B}_i$ are 
\bee\label{pc2}
v_{i\alpha_j}^{\prime}e^{\mu_1\cdots\mu_F;\alpha_1\cdots\alpha_j\cdots\alpha_B}_i=0
\eee
which are due to the fact that polarizors $e_i$ are always projected by  $w^{\alpha}_{i\alpha^\prime}$.
The dimension of $\phi^{\mu_1\cdots\mu_F;\alpha_1\cdots\alpha_B}$ is determined by dimensional analysis
\bee\label{pc3}
d(\phi)=3F+B-1
\eee 
By equating the dimensions of both sides of Eq.(\ref{pc1}), one can derive the minimum of $\tau_i$ 
\bee\label{pc4}
\tau_i^{\min} =2F+B+\frac{1}{2}[1-(-1)^B]\ .
\eee
It is obvious from Eq.(\ref{pc4}) that there are only finite numbers of fermion lines, gluon lines and derivatives contributes to a given power of $1/Q^2$.

\subsection{Collinear Expansion for Arbitrary Loop Orders}
The extension of the collinear expansion for tree diagrams can be straightforward to diagrams containing arbitrary loop corrections. The starting point is to notice that the collinear expansion for the one loop corrections in the collinear region of the radiative gluons can be written down as
\bee\label{cla1}
&&(\sigma_0^{(0)} + \sigma_1^{(0)}+\sigma_0^{(1)} + \sigma_1^{(1)})\Bigg|_{\makebox{\tiny collinear gluons}}\nn
&\approx& \sum_{i=0}^1 \sum_{j=0}^i [(\phi^*_0)^{(i)}\otimes \overline{\sigma}_p^{(0)}\otimes\phi_0^{(i-j)}
+(\phi^*_0)^{(i)}\otimes \overline{\sigma}_p^{(0)}\bullet(\phi_1^H)^{(0)}\otimes \phi_1^{(i-j)}
+(\phi^*_0)^{(i)}\otimes (\phi_1^{*H})^{(0)}\bullet\overline{\sigma}_p^{(0)}\otimes \phi_1^{(i-j)}]\nn
\eee
where superscript $(0),(1)$ denote tree and one loop corrections, repsectively. This is because, as the collinear gluons with momentum $l\sim (Q,\lambda^2/Q,\lambda)$ go through the fermion lines, the valence fermion momenta behave similarly to those in the tree level expansion. This leads to the fact that the collinear expansions for one loop amplitudes in collinear region can be performed just like for tree amplitudes. The soft gluon corrections can not affect the collinear expansion. The cancellations of double logarithms are assured in light-cone gauge by adding ladder and selfenergy diagarms. The one loop corrected parton amplitudes are determined by subtracting the amplitudes in collinear and soft regions from the full one loop amplitudes . Following the standard considerations \cite{twy2}, the LT paron amplitude $\overline{\sigma}_p^{(0)}$ and NLT paron amplitude $\overline{\sigma}_p^{(0)}\bullet(\phi_1^H)^{(0)}$ are infrared finite, and the soft divergences are absorbed by $\phi_0^{(1)}$ and $\phi_1^{(1)}$. The one loop factorization is derived up to NLT order
\bee\label{cla2}
&&(\sigma_0^{(0)} + \sigma_1^{(0)}+\sigma_0^{(1)} + \sigma_1^{(1)})\nn
&\approx&[\sum_{j=0}^1\phi_0^{*(j)}]\otimes [\sum_{i=0}^1 \overline{\sigma}_p^{(i)}]\otimes[\sum_{j=0}^1\phi_0^{(j)}]\nn
&&+[\sum_{j=0}^1\phi_0^{*(j)}]\otimes [\sum_{i=0}^1 \sum_{j=0}^i\overline{\sigma}_p^{(j)}\bullet(\phi_1^H)^{(i-j)}]\otimes [\sum_{k=0,1} \phi_1^{(k)}]\nn 
&&+[\sum_{j=0}^1\phi_0^{*(j)}]\otimes [\sum_{i=0}^1 \sum_{j=0}^i(\phi_1^{*H})^{(i-j)}\bullet\overline{\sigma}_p^{(j)}]\otimes [\sum_{k=0,1} \phi_1^{(k)}]\ .
\eee
The generalization to arbitrary loop orders can be obtained by iteration.
Suppose that 
\bee
\sigma=\phi_0^*\otimes(\sigma_p)_0\otimes\phi_0+\phi_0^*\otimes(\sigma_p)_1\otimes\phi_1+\phi_1^*\otimes(\sigma_p)_1^*\otimes\phi_0
\eee
where
\bee
\sigma&=&\sum_{i=0}^N \sigma^{(i)}\ , 
(\sigma_p)_0=\sum_{i=0}^N (\overline{\sigma}_p)_0^{(i)}\ ,
\phi_{0,1}=\sum_{i=0}^N \phi_{0,1}^{(i)}\nn
(\sigma_p)_1&=&\sum_{i=0}^N (\sigma_p)_1^{(i)}\ ,
\eee
where
\bee
(\sigma_p)_1^{(i)}=\sum_{j=0}^N \overline{\sigma}_p^{(j)}\bullet(\phi_1^H)^{(i-j)}\ .
\eee
The above factorization still holds for $N+1$ order corrections, since the collinear gluons cannot attach to the parton amplitudes. The remaining proof of factorization requires the cancellations of double logarithms of soft divergences, the single soft logarithms absorbed by pion DAs and the infradred finitness of the parton amplitudes. This can be achieved by standard analysis (see e.g. \cite{Lepage:1980fj}) and it is left to other publish \cite{twy2}.

\section{$O(Q^-4)$ contributions }
The leading twist contribution of the pion form factor $F_\pi(Q^2)$ is expressed as
\begin{eqnarray}\label{ltf1}
F_\pi^{(t=2)}(Q^2)=\frac{128\pi\alpha_s(Q^2_{\makebox{\small eff}})}{9\bar{Q}^2} \int dx_1 dx_2 \frac{\phi(x_2)\phi(x_1)}{x_1 x_2}\ ,
\end{eqnarray}
where $\bar{Q}^2=(1-\cos\theta)Q^2/2$ has been employed . Applying   $\phi(x_i)=3f_\pi x_i(1-x_i)/\sqrt{2}$ into Eq.~(\ref{ltf1}), one can get
\begin{eqnarray}\label{ltf2}
F_\pi^{(t=2)}(Q^2)=\frac{16\pi\alpha_s(Q^2_{\makebox{\small eff}})f_\pi^2}{\bar{Q}^2}\ .
\end{eqnarray}
We use $f_\pi=93$ MeV in this paper.

The Feynman diagrams displayed in Fig.2 contribute to $O(Q^{-4})$. The calculation is straightforward and the result reads
\begin{eqnarray}
F_{\pi}^{(t=4)}(Q^2)=-\frac{256\pi\alpha_s(Q^2_{\makebox{\small eff}})}{9\tilde{Q}^4}\int dx_1 dx_2 \phi(x_2)\frac{[G(x_1)+\tilde{G}(x_1)(1-2x_1)]}{x_2 x^2_1(1-x_1)}+(x_1\longleftrightarrow x_2)\ ,
\end{eqnarray}
where $G$ and $\tilde{G}$ mean the twist-4 pion distribution amplitudes \cite{Yeh:2001gu}.
Note that we have employed the effective coupling $\alpha_s(Q^2_{\makebox{\small eff}})$ with argument $Q^2_{\makebox{\small eff}}\equiv <x_1 x_2 Q^2>$, where the bracket means the average under the nonperturbative QCD vaccum. This is contrary to the conventional treatment that $\alpha_s(Q^2)$ is running with $Q^2$. The scale $x_1 x_2 Q^2$ of $\alpha_s$ is from the momentum of the exchange gluon. 
From the factorization point of view, we may write $<x_1 x_2 Q^2>$ into $<x_1><x_2> Q^2_{\makebox{\small fact}}$ by setting a factorization scale $Q_{\makebox{\small fact}}$ to marker the border between the perturbative and the nonperturbative dynamics. 
The scattering angle dependence has been absorbed by $\tilde{Q}$ as $\tilde{Q}^4=Q^4(1-\cos\theta)^2/2(1+\cos\theta)$. The dependence in $\theta$ is irrelevant for power correction, because it is hard to be identified by experiment. The contribution from $\tilde{G}(x_i)$ is suppressed by the factor $(1-2 x_i)$. Therefore, the contribution mainly comes from $G(x_1)=3\sqrt{2}\pi^2 f_\pi^3 x_1(1-x_1)$ \cite{Yeh:2001gu}. After substituting the distribution amplitudes and performing the integrations, there remaines an infrared divergence \cite{Geshkenbein:1982zs,Li:2001ay}
\begin{eqnarray}
F_{\pi}^{(t=4)}(Q^2)=-\frac{256\pi\alpha_s(Q^2_{\makebox{\small eff}})\pi^2 f_\pi^3}{\tilde{Q}^4}\int \frac{dx_1}{x_1}\ .
\end{eqnarray}
It can not be completely resolved under perturbation theory. There requires a resummation over the soft divergences as the virtual quark lines become on-shell. One also needs to introduce a jet function to absorb these divergences. We will skip the details of the perturbative behavior of these divergences \cite{Geshkenbein:1982zs,Geshkenbein:1982zu,Geshkenbein:1984qn,Li:2001ay}. The initial function for such a jet function is of nonperturbative. We denote the effective $J$ function as $J(Q^2_{\makebox{\small eff}})$ and rewrite the NLO form factor as
\bee
F_{\pi}^{(t=4)}(Q^2)=-\frac{256\pi\alpha_s(Q^2_{\makebox{\small eff}})\pi^2 f_\pi^3 J(Q^2_{\makebox{\small eff}})}{\tilde{Q}^4}\ .
\eee
Adding $F_{\pi}^{(t=2)}(Q^2)$ and $F_{\pi}^{(t=4)}(Q^2)$, we obtain 
\bee
F_{\pi}(Q^2)=\frac{16\pi\alpha_s(Q^2_{\makebox{\small eff}})f^2_\pi}{Q^2}(1-\frac{16\pi^2f^2_\pi J(Q^2_{\makebox{\small eff}})}{Q^2})+O(Q^{-6})\ .
\eee

Inspired by the theoretical result, we can perform a least $\chi^2$ ($\chi^2_{\min} =7.96742$) fit for the data \cite{Bebek:1976wv,Bebek:1978pe,Amendolia:1986wj} to obtain 
\bee
F_\pi^{\makebox{\small Fit}}(Q^2)=\frac{0.46895}{Q^2}(1-\frac{0.3009}{Q^2})\ .
\eee
The $\chi^2$ analysis for the data is shown in Fig.3. It is obvious that the data point at $10$ GeV$^2$ is out of allowed errors. By igoring the 10 GeV$^2$ data point, we can find that the $Q^2\to\infty$ limit of $Q^2 F_\pi(Q^2)$ approaches a constant, supporting our taking $\alpha_s(Q^2_{\makebox{\small eff}})$ in the above calculations.
Comparing the fit formula and the theory formula for $F_\pi(Q^2)$, we are led to the following conclusions:
\begin{enumerate}
\item The argument $Q^2$ of $\alpha_s(Q^2)$ should be interpreted as an effective $Q^2_{\makebox{\small eff}}\equiv <x_1><x_2> Q^2_{\makebox{\small fact}}$. That is we need to take $\alpha_s(Q^2)$ as an effective coupling constant. The average fractions $<x_1>\approx <x_2>\approx 0.57$ and the factorization scale $Q^2_{\makebox{\small fact}}=1$ GeV for $\Lambda_{\makebox{\small QCD}}=0.3$ GeV. The values of $<x_i>$ and $Q^2_{\makebox{\small fact}}$ depend on the model of the pion wave function we have employed (the AS model). If we perform similar analysis by employing CZ model $\phi(x)=15f_\pi x(1-x)(1-2x)^2/\sqrt{2}$ and $G(x)=15\sqrt{2}\pi^2f_\pi^3x(1-x)(1-2x)^2$ \cite{Yeh:2001gu}, we may then obtain $Q^2_{\makebox{\small fact}}=13.12$ GeV$^2$ for $<x_i>\approx 0.5$ or $Q^2_{\makebox{\small fact}}=328$ GeV$^2$ for $<x_i>\approx 0.1$. It is clear that CZ model is less consistent with PQCD than AS model. 
\item A less model dependent property of the effective coupling constant can be derscribed: The change in $\Lambda_{\makebox{\small QCD}}$ would affect the location of the average fraction variable $<x>$ for a fixed factorization scale $Q_{\makebox{\small eff}}$. On the other hand, for a fixed $<x>$, $Q_{\makebox{\small eff}}$ would vary with $\Lambda_{\makebox{\small QCD}}$. Nevertheless, there are only finite possible consistent solutions for $\Lambda_{\makebox{\small QCD}}$,  $Q_{\makebox{\small eff}}$ and $<x>$ can be derived. 
\item The effective value of $J$ function at $Q^2_{\makebox{\small eff}}$ is about 0.22. 
\item If we combine the result $F_{\gamma\pi}(Q^2)$ of \cite{Yeh:2001gu}
\bee
F_{\gamma\pi}(Q^2)=\frac{2f_\pi}{Q^2}(1-\frac{8\pi^2 f_\pi^2}{Q^2})\ ,
\eee
we may extract an effective running coupling constant \cite{Lepage:1980fj} 
\bee
\alpha_s^{\makebox{\small eff}}(Q^2/4)=\frac{F_\pi^{\makebox{\small Fit}}(Q^2)}{4\pi Q^2 F_{\gamma\pi}^2(Q^2)}\ .
\eee
The factor one fourth in $\alpha_s^{\makebox{\small eff}}(Q^2/4)$ is from $<x_i>\approx 0.5$. The comparison between $\alpha_s^{\makebox{\small eff}}(Q^2)$ and the usual running coupling constant $\alpha_s(Q^2)$ with $\Lambda_{\makebox{\small QCD}}=0.3$ GeV is shown in Fig.4. One can see that their difference is very small for all range $Q^2 > 1$ GeV$^2$. To reduce their differences, higher order power corrections are required. 
The effective running coupling constant $\alpha_s^{\makebox{\small eff}}/\pi$ is smaller than 0.2 for $Q^2>1$ GeV$^2$.  At the accuarcy we work, it directly showes that PQCD is applicable to exclusive processes for momentum transfer of $Q^2>1$ GeV$^2$ ($\alpha_s/\pi<0.5$).
\end{enumerate}

\section{Discussions}
From the above analysis, we may draw the conclusion that PQCD is applicable to $\gamma^*\pi\to\pi$ at the energy region accessable by the experiment. In our analysis, we only employed the simplest models without invoking any nonperturbative arguements. From our result, we may try to understand the evidence showed by Isgure and Llewenlly Smith.  In their analysis, they employed the fraction function
\bee
f(\epsilon)=\int_0^1 dx \int_0^1 dy \theta(xy - \epsilon)\frac{\phi(x)\phi(y)}{xy}
\eee
to evaluate the percentage of perturbation contributions. The parameter $\epsilon$ describes a cut-off on $xy$ in order to keep higher-twist contributions and higher-order effects at a reasonable small level. Since $f(\epsilon)$ is a mixing of perturbative and nonperturbative contributions, the perturbation parts are those corresponding to large values of $xy Q^2$.
For the AS model, $f(\epsilon)$ reaches 90 $\%$ for $\epsilon=1/150$. It implies that the naive perturbative contribution is $90\%$ legal for $Q^2=150$ GeV$^2$. It is noted that they employed a common setting for the scale of the running coupling constant $\alpha_s(Q^2)$ in the pion factor factor, the small value of $\alpha_s(Q^2)$ for $Q^2\ge 1$ GeV$^2$ leads to their conclusion that PQCD is inapplicable to exclusive process.

However, the real value of the coupling constant or the factorization scale  involving in process $\gamma^*\pi\to\pi$ can not be determined a priori by perturbation theory. On the contrary, it should be determined from other methods, e.g. the experiment. In fact,  $f(\epsilon)$ is reasonable only for the asymptotic region at $Q^2\to\infty$. As shown above, it is indeed. But the lack of experiment in such a high $Q^2$ region, it is difficult to determine the factorization scale. The analysis of Isgur and Llewellyn Smith employed the leading result, which should be valid for $Q^2\to\infty$. Therefore, they needed to extrapolate the leading result from high $Q^2$ to low $Q^2$. It seems that the conclusion of Isgur and Llewellyn Smith is weak. Because we have employed the leading and sub-leading contributions in the pion form factor, we can control the $Q^2$ behaviors of pion form factor for both high and low $Q^2$. Therefore, our conclusion is stronger and more reliable.

The power corrections to the pion form factor have also been calculated by employing the projection for the pion \cite{Geshkenbein:1982zs,Geshkenbein:1982zu,Geshkenbein:1984qn,Cao:1999st}
\bee
P(k,p-k)=\frac{1}{4}\gamma_5\{1+\frac{2(2x-1)}{Q^2}p_{\mu}\sigma^{\mu\nu}p_{\nu}^{\prime}+\frac{2x(1-x)}{k_{\perp}^2}p_{\mu}\sigma^{\mu\nu}k_{\perp\nu}\}
\eee
where $k$ is the momentum carried by the valence quark of the pion, $x$ represents the fraction $x=k\cdot p^{\prime}/p\cdot p^{\prime}$ and $p$ and $p^{\prime}$ denote the momenta carried by the initial and final state pions. The result appears as
\bee
F_\pi(Q^2)=\frac{16\pi\alpha_s(Q^2)f_{\pi}^2}{Q^2}\{1+\frac{m_{\pi}^4}{Q^2 m_0^2}J^2(Q^2)\}\ .
\eee
The function $J(Q^2)$ represents the $J$ function introduced before. The factor $\frac{m^2_{\pi}}{m_0}\ge 1.4$ GeV results in a violation of perturbation principle for the region of $Q^2$ being $2\sim 15$ GeV$^2$. The positive sign in $O(Q^{-4})$ power contributions seems difficult to expain the $\chi^2$ fit formula of the data.

Another factor that would affect the leading result may arise from the contributions of Sudakov form factors. However, as shown in \cite{Lepage:1980fj} that the Sudakov form factor falls fastly than any power suppression for low $Q^2$. Then, the low $Q^2$ behaviors of the pion form factor is mainly controlled by the power corrections (the sub-leading contributions).

\section{Conclusions}
We have developed a power expansion scheme for $\gamma^*\pi\to\pi$. This expansion scheme also has been shown to be compatible with QCD factorization. The NLO power corrections to the pion form factor have been calculated. With the help of the formula of the pion form factor upto $Q^{-4}$, we may perform a least $\chi^2$ analysis for the data. By comparing the $\chi^2$ fitting formula and the theory formula of the pion form factor, we arrive at the conclusion that, due to the nonperturbative QCD vaccum, the strong coupling constant in the theory formula should be identified as an effective coupling with factorization scale equal to 1 GeV. In addition, the averaged fraction variable locates at 0.5 in consistency with the AS model for pion wave function. The CZ model is fail to give a consistent explanantion for the data, because its relative factorization scale is in the range of $3\sim 18$ GeV. 

We also derive the $Q^2$ behavior of the effective coupling, the effective running coupling constant. From the $Q^2$ analysis of the effective running coupling constant, we may prove that PQCD is applicable to exclusive process for momentum transfer larger than 1 GeV$^2$. This is a direct proof showing that PQCD is applicable to exclusive process for $Q^2>1$ GeV$^2$. In addtion, the analysis performed by Isgur and Llewellyn Smith showing that PQCD is inapplicable to exclusive process, can also be undertood that the leading contribution is valid in the asymptotic region and can not be applied for low $Q^2$.

The result obtained in this paper can be easily extended to other hard processes involving two light mesons, such as $\eta,\rho\to\pi\gamma$, etc. 

From the result of this paper and \cite{Yeh:2001gu}, we have a suggestion that the power corrections in exclusive processes are sizable and require detail investigations.  Over past decades, we have accumulated abundant data for exclusive processes.  Most data are in low energy region. This is the place where the power correction become important and can not be negligible. The leading or the asymptotic contribution only gives information about the hadron wave function, the static property of QCD, while the power correction can reveal the dynamics. This is why the power correction plays an important role in analysis of data.
%%%%%%%%%%%%%%%%%%%%%%%%%%%%%%%%%%%%%%%%%%%%%%%%%%%%%%%%%%%%%%%%%%%%%%%%%
\vspace{0.5cm}

\noindent
{\bf Acknowledgments:}
This work was supported in part by the National Science Council of R.O.C. under the Grant No. NSC89-2811-M-009-0024.

Figure Caption
\begin{list}{}
\item Fig.1 The leading order diagrams for $\gamma^*\pi\to\pi$. 
\item Fig.2 The next-to-leading-twist (NLT) diagrams for $\gamma^*\pi\to\pi$. The propagator with one bar means the special propagator.
\item Fig.3 Plot of the least $\chi^2$ fit (solid line) and C.L.$=99.73\%$ (dash line) for $Q^2 F_{\pi}(Q^2)$. The experimental data are taken from \cite{Bebek:1976wv,Bebek:1978pe,Amendolia:1986wj}.
\item Fig.4 The comparison between the effective running coupling constant (solid line) and the perturbative running coupling constant (dash line).
\end{list}

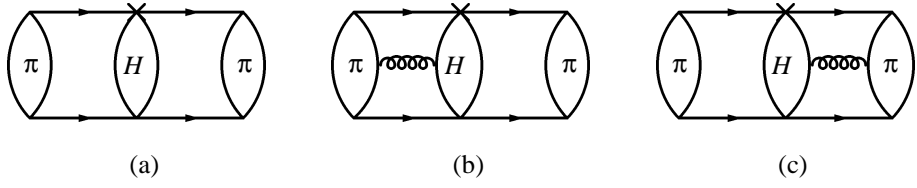
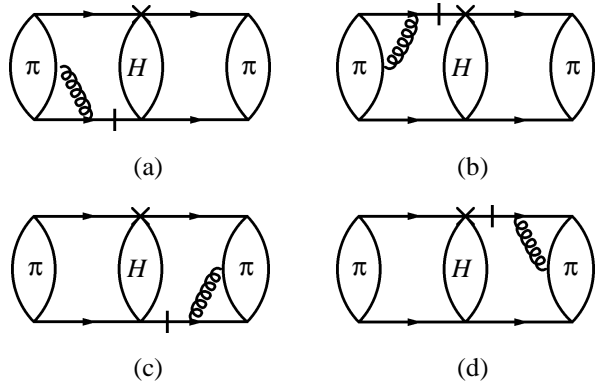
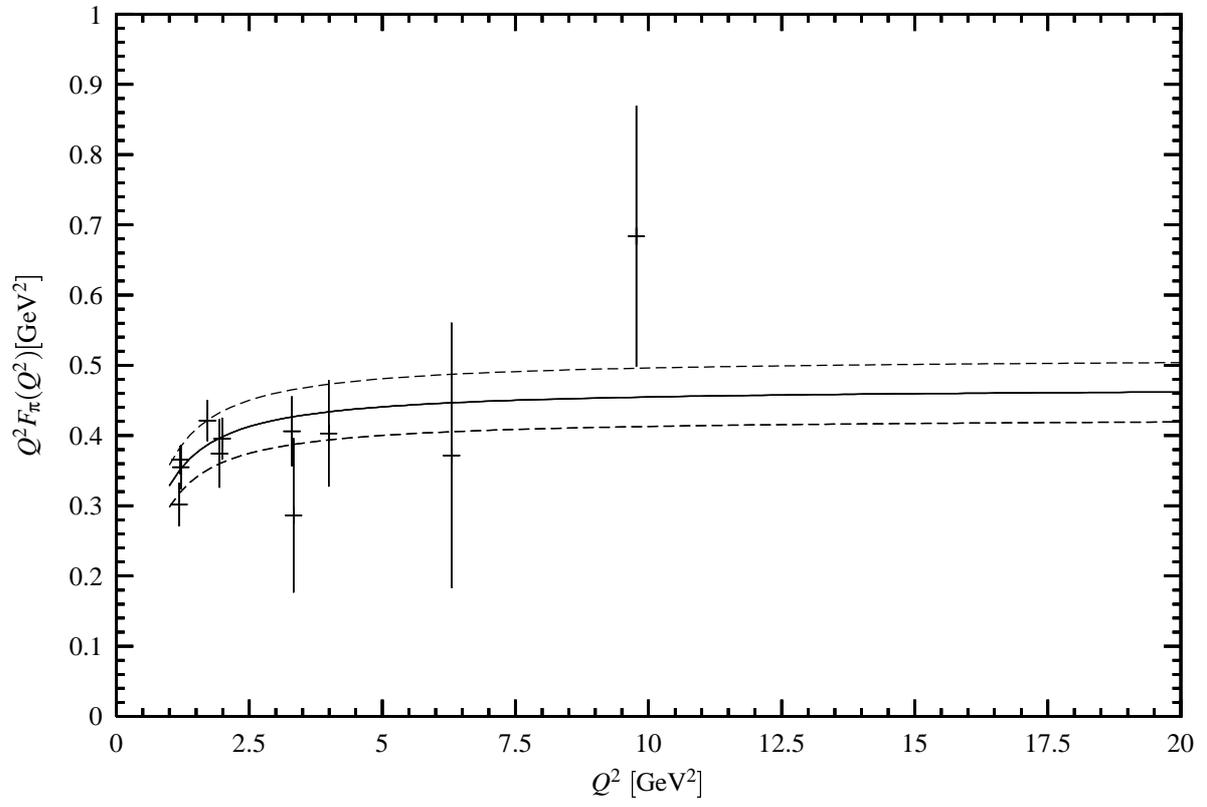
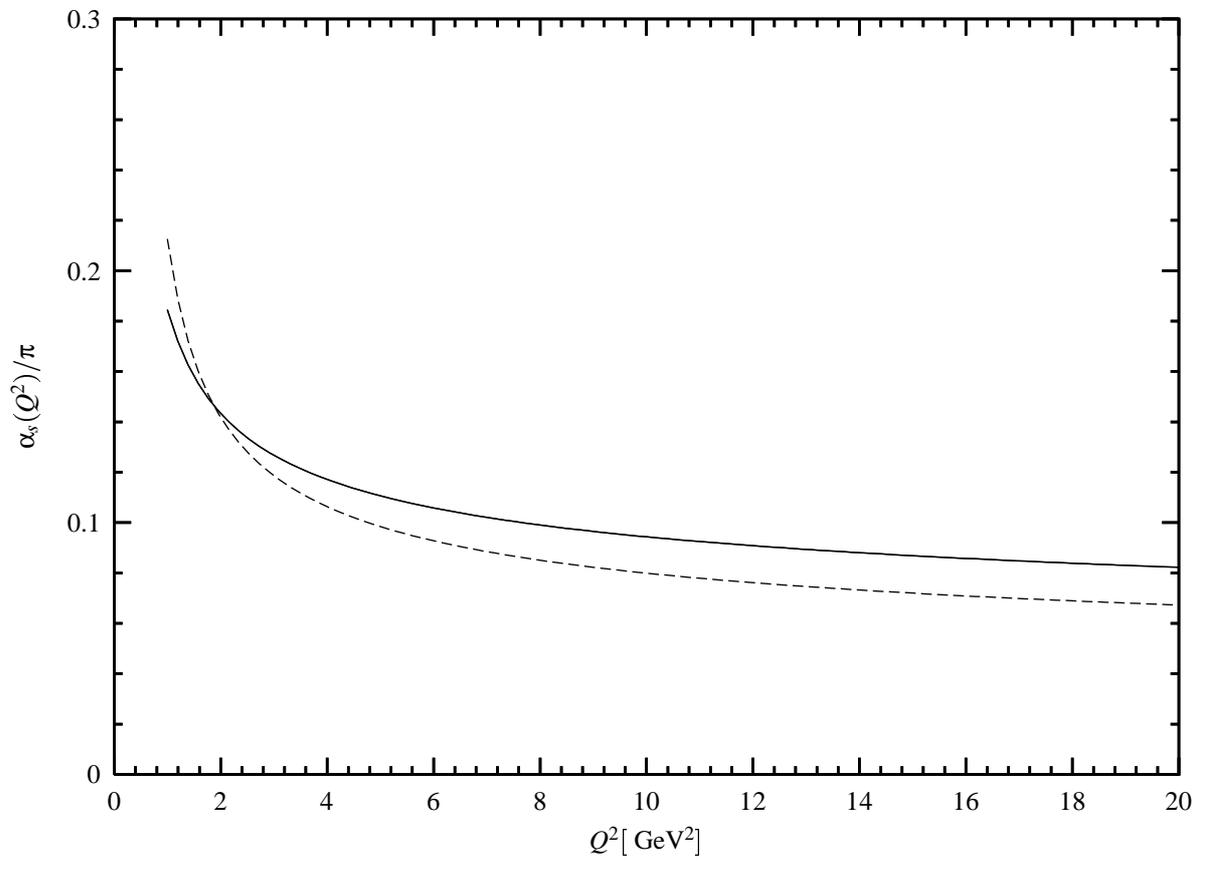
\end{document}